\documentclass[reprint,amsmath,amssymb,aps,]{revtex4-1}

\usepackage{graphicx}
\usepackage{dcolumn}
\usepackage{bm}

\begin{document}
\preprint{APS/123-QED}

\title{A road map for synthesizing the scaling patterns in ecology}

\author{Cang Hui}

\affiliation{Centre for Invasion Biology, Department of Botany and Zoology, Stellenbosch University, Matieland 7602, South Africa; E-mail: chui@sun.ac.za}

\date{\today}

\begin{abstract}
Ecology studies biodiversity in its variety and complexity. It describes how species distribute and perform in response to environmental changes. Ecological processes and structures are highly complex and adaptive. In order to quantify emerging ecological patterns and investigate their hidden mechanisms, we need to rely on the simplicity of mathematical language. This becomes especially apparent when dealing with scaling patterns in ecology. Indeed, nearly all of ecological patterns are scale dependent. Such scale dependence hampers our predictive power and creates problems in our inference. This challenge calls for a clear and fundamental understanding of how and why ecological patterns change across scales. As Simon Levin stated in his MacArthur Award lecture, the problem of relating phenomena across scales is the central problem in ecology and other natural sciences. It has become clear that there is currently a drive in ecology and complexity science to develop new quantitative approaches that are suitable for analysing and forecasting patterns of ecological systems. Here I provide a road map for future works on synthesizing the scaling patterns in ecology, aiming (i) to collect and sort a diverse array of ecological patterns, (ii) to present the dominant parametric forms of how these patterns change across spatial and temporal scales, (iii) to detect the processes and mechanisms using mathematical models, and finally (iv) to probe the physical meaning of these scaling patterns. This road map is divided into three parts and covers three main concepts of scale in ecology: heterogeneity, hierarchy and size. Using scale as a thread, this road map and its following works weave the kaleidoscope of ecological scaling patterns into a cohesive whole.

\begin{description}
\item[PACS numbers]
87.23
\end{description}
\end{abstract}
\pacs{Valid PACS appear here}
\maketitle

Ecological patterns are emerging structures observed in populations, communities and ecosystems. Elucidating drivers behind ecological patterns can greatly improve our knowledge on how ecosystems assemble, function and respond to change and perturbation. Due to the non-random nature, most, if not all, ecological patterns change with measurement, characteristic and organization scales and exhibit distinct scaling properties. Such scaling properties can be broadly grouped into patterns related to heterogeneity, hierarchy and size. The road map introduces the three groups of scaling patterns. The emphasis here is not to provide solutions to these outlined research questions; rather, by grouping relevant scaling patterns under unique banners, I attempt to highlight the challenges and connect the emerging clues for building a unified theory for scaling patterns in ecology in the near future.

\section{Heterogeneity}
This section on the scaling pattern of heterogeneity aims to investigate how aggregated structures of organisms, diversity and ecosystem service change with measurement scales and which/why biological patterns resonate with underlying processes at the same characteristic scales.

\textbf{Aggregation.}
Species distributions are not uniform across space, reflecting the interplay between habitat heterogeneity and the underlying nonlinear biotic regulation \cite{ref1}. Such non-random, aggregated patterns not only can be the indicative of non-equilibrium dynamics (e.g. during range expansion \cite{mihaja,ecography,cecile}) but also self-organized pattern emergence (e.g. \cite{ploszhang,ncer,ncem}). When ecologists examine species distributions across scales, the Modifiable Areal Unit Problem presents itself \cite{ref2,maup}. The problem can be described as the change in species distribution characteristics as the unit of measurement changes, both in terms of size and shape of the sample unit. Such scaling patterns of aggregation follow three general parametric forms: logarithmic, power-law and lognormal shape \cite{ref1}. Following on recent progress of using the Bayesian rule for cross-scale extrapolation \cite{ref3,ref4}, further advancement in this field is to provide a consistent description of aggregation when scaling up and down, and therefore a universal basis of comparison for distributions in differing contexts. A fully-functional model with predictive power for up- and down-scaling species distribution is needed. Under certain conditions, this model should further allow extrapolating fine-scale occupancy and population densities from coarse-scale observations (e.g. \cite{ecoappl,ecoscience}). Great potential exists to apply such a predictive model in various cross-scale pattern analyses espcially when detection is imperfect (e.g. \cite{springer,jae}).

\textbf{Space-for-time substitution.}
The directionality of community succession is an important concept in conservation biology \cite{Odum,pickett}; it is analogous to the irreversibility of time in physics that has revolutionised the understanding of complex adaptive systems \cite{jorg}. By definition, succession is {\em an orderly process of community change} after disturbance \cite{Odum}. Knowing the directionality of succession is necessary for (i) distinguishing new from mature communities (i.e. defining the age of a community), (ii) understanding how communities evolve and respond to disturbance (e.g. habitat loss and climate change), and (iii) designing more efficient conservation and restoration plans \cite{nuria}. However, popularising the concept of directionality in succession is challenging for two reasons. Firstly, acknowledging this concept demands the acceptance of inherent bias in nature which contradicts the null hypothesis of a random and isotropic world \cite{jorg}. Secondly, appropriate long-term data required for detecting the succession direction are scarce, and indices and analytical methods for such computation are lacking. To this end designing alternative tests (e.g. the space-for-time substitution; \cite{pickett} that can capture the essence of directionality and irreversibility in community development but which can be applied to available data becomes crucial.

The spatial and temporal scales of ecological processes are intertwined. Processes that account for the spatial distribution of species also underpin its temporal dynamics. This means that we can potentially forecast the future or rebuild the history based solely on current spatial distribution, without resorting to long-term time series \cite{ref5,ref6}. In other words, the need to wait years and decades to measure changes in distribution can be averted through the ability to make sufficiently accurate predictions based only on the spatial distribution of species at the current time. As the ability to forecast the temporal trend of a focal species provides crucial information on its performance and viability \cite{ref5,ref7}, the methodology of space-for-time substitution is extremely appealing, especially because our ability to obtain spatial records has been drastically improved. This area of research calls for a model that can relate the scaling pattern of species’ current spatial distribution to the near-future population trend and performance.

\textbf{Scale resonance.}
Just as two tuning forks of the same characteristic frequency resonate, so do ecological patterns and processes working at the same scale. Species distributions are regulated by a variety of abiotic and biotic processes working in concert but at different scales \cite{mcgill}. Those processes identified as key biotic drivers using methods such as multivariate statistics often resonate with the scale of the study. That is, information being picked up represents a product of the measurement method, rather than the intrinsic cross-scale mechanism. Such a pattern of scale resonance has been observed when synthesizing a series of collaborative works on identifying the factors of the distribution of Argentine ants at local \cite{ref8}, regional \cite{ref9} and global scales \cite{ref10}. This finding brings into question many regional management planning practices that are based on the upscale extrapolation of local-scale studies. Future research needs to explore the mechanism behind scale resonance in ecology and to present a statistical remedy for cross-scale inference.

\textbf{Co-distribution.}
To exploit resources while mitigating conflict, species often partition available habitats, forming co-distribution patterns of association or dissociation. Null models based on permutation have been widely applied for detecting signals of association or dissociation from co-distribution patterns, from which the type of biotic interactions can be inferred. Future research needs to present a model that incorporates biotic interactions and also captures the transition from fine-scale dissociation to coarse-scale association (e.g. \cite{ref4}), explaining why this co-distribution pattern changes across scales and how this scale dependence affects the pairwise measure of species turnover. It should be reconcile the debate between the Rich-Get-Richer phenomenon in invasion biology and the opposing Competitive-Exclusion-Principle.

\textbf{Biodiversity.}
Species diversity patterns, such as the species-area curve \cite{ref11a}, endemics-area relationship, distance decay of similarity and occupancy frequency distribution \cite{ref11,ref12}, are just a few interrelated patterns of scale dependence emerging from complex ecological systems. The integration of patterns of species diversity patterns is central to understanding the processes that drive species assembly \cite{125}. Changing the measurement scale will lead to a coordinated change in all diversity patterns. I envisage  a new diversity pattern – delta diversity – that connects all commonly known diversity scaling patterns, using delta diversity as building blocks. This model should be able to further explain Raunkiaer’s bimodal law of frequency \cite{ref13} and resolves the debate on the ceiling of species richness in a community.

\textbf{Ecosystem function and service}
Ecosystem services are by-products from the function of ecosystem processes that sustain the basic needs of humans and their socioeconomic activities. According to the Millennium Ecosystem Assessment, ecosystem services are generated from interactions ranging from specialist taxa to all biodiversity, and the functional units of the variety of services range from local populations to global biogeochemical cycles. At the local scale, we benefit from services of pollination, pest control, soil fertility and seed dispersal that are related to biodiversity. At regional scale, we benefit from services of air and water purification, flood and drought mitigation, and waste decomposition that are delivered by plants and micro-organisms. At the global level, we benefit from services of climate stability and UV protection from plants and biogeochemical cycles. Understanding how different ecosystem services change with spatial scales and potentially conceptualizing into a model for extrapolating the level of service across scales warrants great attention. Recent  studies showed the scale dependence of a crude ecosystem service indicator – biocapacity and the resultant sustainability index \cite{ref14,ref15}, and the robustness and invasibility of recipient ecosystems to biological invasions \cite{ref16}. Exploiting ecosystem services within their maximum sustainable level can ensure a reliable service provision without triggering a regime shift. Achieving this balance is a challenge for conservation management and sustainable development.

\section{Hierarchy}
The scaling pattern of hierarchy depicts how the structure and function of asymmetrical ecological systems emerge and change with the system complexity. Using ecological networks as the model system, this section aims to investigate how cascade interactions affects the robustness and resilience of networks, how network architectures, especially nestedness and compartmentalization, emerge and function, and the role of network complexity on the stability of ecological networks.

\textbf{Cascade.}
Nodes and edges of a network are a good proxy of species and their interactions in an ecosystem. Probing processes that can lead to the emergence of large-scale network architectures, e.g. small-world networks and scale-free networks, is a new wave in science. Scaling laws of food webs depict how a biological network behaves as a function of its complexity. In reconciling with May’s stability criterion of complex systems \cite{may}, Cohen’s cascade model \cite{cohen} and its later development provide a phenomenological explanation of some of the scaling laws and scaling invariant patterns. The principle of Maximum Entropy that identifies the unbiased estimates under constraints has been widely used in ecology. Ecological networks are efficient energy transporting, non-equilibrium systems that are adaptive to changes and disturbances. This calls for the development of models based on the recently developed principle of Maximum Entropy Production in non-equilibrium thermodynamics \cite{dewar}, in an attempt to explain the pattern emergence in complex adaptive networks, in particular, the cascade interactions in food webs. Such models  will provide a physical understanding on the network emergence and shed light on solving the complexity-stability debate.

\textbf{Nestedness.}
Nested structure has been observed in many networks, in particular bipartite mutualistic networks (e.g. pollination networks and seed-dispersal networks \cite{bas}). To have models with quantitative accuracy and predictive power, one needs to rely on process-based models. A key feature in ecological networks is the adaptive and innovative nature of the edges and nodes. Species are constantly optimizing which partners they should interact with for maximum fitness gain, as according to the optimal foraging theory. This can be achieved at two time scales: 1) At the ecological time scale, interaction switching reflects seeking optimal fitness gain \cite{ref17,anim}; 2) At the evolutionary time scale, interaction switching reflects coevolutionary dynamics between interacting traits.

\textbf{Compartmentalization.}
One important hierarchical structure in ecological networks is compartmentalization, i.e. the formation of functional modules, where interactions are most likely to occur within the same module, that is to say “like is connected to like” in a network \cite{ref18}. Energy flows directionally in resource-exploitation networks (e.g. host-parasite networks), forming compartmentalized structures. This research area calls for a process-based model that optimizes energy transport via adaptive interaction switching partners for local fitness gain. The strong predictive power requires further demonstration by comparing the level of compartmentalization observed versus predicted for real networks \cite{savannah}. This, together with the previous two research areas, provides a physical understanding of pattern emergence in complex adaptive networks.

\section{Size}
This section on the scaling pattern of size investigates how form and function change as organisms get larger or as their traits change. The section aims to investigate how allometric laws of metabolism emerge and how phenotypes of biological traits form through co-evolution with other traits, species and the environment and how this affects the path of evolution and diversification.

\textbf{Allometry.}
Allometric scaling is the most salient pattern of how biological rates, especially metabolic rate, are regulated by organism size \cite{allo}, known as Kleiber’s law. A model, depicting how fractal-structured circulatory or vascular systems transport energy and matter throughout the body of the organism, has been proposed for explaining the emergence of allometric scaling, which is further summarized succinctly as the Fourth Dimension of Life \cite{west}. Recently, a not fully developed concept suggests that natural selection can increase the ontogenetic and population growth rates to a ceiling, above which the population will crash due to intrinsic instability \cite{lev}, in principle similar to the origin of planetary rings that are only distributed on a thin surface surrounding the planet, with all other directions, although possible, have been eliminated through inter-particle collisions. This progress calls for further development of a conceptual model for allometric scaling based on the principle of Maximum Entropy Production and self-organized criticality. This model should be able to simulate the process of cell differentiation in multicellular organisms.

\textbf{Trait.}
This research area focuses on addressing two questions. First, what are the mechanisms determining phenotypic traits? Evolution via natural selection relies on heritable phenotypic variation and has long been regarded as being solely reliant on direct expression of gene variation. This assumption may be an exaggeration, as factors other than genetic variation can also dictate the outcomes of natural selection and thus evolution. At the forefront of these alternative platforms is the rapidly expanding field of epigenetics – the biochemical modification of DNA without changes in its sequence – that gives rise to differential gene expression and thus different phenotypes \cite{epi}. A revised Price equation that incorporates epigenetic mechanism for explaining phenotypic variation is needed. Second, how do traits affect biotic interactions and thus the path of evolution? Biotic interactions, largely to do with resource competition and exploitation, are realised by specific phenotypic traits that are used for searching and handling resources, for competing and defending resources, for memorizing and recognizing beneficial resources, and for escaping from being exploited as resources by predators or parasites. In this regard, Adaptive Dynamics is a power tool to explore how trait-mediated interactions can lead to diverse evolutionary phenomena \cite{ulf}, from the Red Queen dynamics to speciation \cite{ec,ref19}.

\section{Epilogue}
Nature never fails to amaze us. It coordinates the interplay of numerous organisms in their environments, forming a complex functional system that sustains us and many other species via ecosystem service. As the ultimate goal of natural sciences, quantifying emerging patterns in nature and understanding hidden mechanisms are the pinnacle of science. An important phenomenon in quantifying ecological patterns is that nearly all of them are scale dependent. This scale dependence creates problems in our inference, yet also simultaneously provides opportunities for us to pry into the core of how nature assembles, organizes and functions. Once again, we need to relate these research areas back to the overarching research philosophy in studying natural systems: (i) what patterns exist in nature (i.e. using statistical methods to measure and quantify ecological patterns); (ii) how such patterns emerge (i.e. proposing mathematical models to unveil mechanisms driving the ecological pattern formation; (iii) why nature organizes itself in such a way (i.e. using physical laws to reveal the adaptive and/or optimal nature of ecological systems). This has been eloquently presented as the Ouroboros of scales in {\em Just Six Numbers} by Martin Rees. William Blake also expressed the role of scales in his {\em Auguries of Innocence} in 1803, `To see a world in a grain of sand, and a heaven in a wild flower; Hold infinity in the palm of your hand, and eternity in an hour'.

\end{document}